\documentclass[runningheads,a4paper]{llncs}
\usepackage{graphicx}
\usepackage[usenames,dvipsnames]{color}
\usepackage{float}
\usepackage{placeins}
\usepackage{url}
\usepackage{listings}
\usepackage{enumitem}
\setlist{nolistsep}

\usepackage{amsmath}
\usepackage{textcomp} % \textlbrackdbl, \textrbrackdbl
\usepackage{xspace} 
\usepackage{bbding} % check marks
\usepackage{pifont} % arrows
\usepackage{stmaryrd} % arrows
\usepackage{mdwlist} %list 

\newcommand{\Focus}{\textsc{Focus}}
\newcommand{\epar}[1]{``{#1}"}

\begin{document}

\title{ Model-based generation\\
 of natural language specifications}

\author{Phan Vo Thu Nhat, Maria Spichkova}
\institute{RMIT University, Melbourne, Australia \\  
\email{s3220976@student.rmit.edu.au, maria.spichkova@rmit.edu.au} }
 
\maketitle

\begin{abstract}  
Application of formal models provides many benefits for the software and system development, however, the learning curve of formal languages could be a critical factor for an industrial project. Thus, a natural language specification that reflects all the aspects of the formal model might help to understand the model and be especially useful for the stakeholders who do not know the corresponding formal language. 
Moreover, an \emph{automated generation} of the documentation from the model would replace manual updates of the documentation for the cases the model is modified.
This paper presents an ongoing work on  generating natural language specifications from formal models. 
Our goal is to generate documentation in English from the basic modelling artefacts, such as data types, state machines, and architectural components.
To allow further formal analysis of the generated specification, we restrict English to its subset, Attempto Controlled English. \footnote{%
Preprint. Accepted to the Software Technologies: Applications and Foundations (STAF 2016). Final version published by Springer International Publishing AG.
}
\end{abstract}

%========================================
\section{Introduction}

Model-based development (MBD) is a paradigm in which software and system development focus on high-level executable models, cf. \cite{zhang2006model}. 
In the early development phases, formal models allow a wide range of exploration and analysis using domain-specific notations in order to simplify the system design, development or verification/testing. 
Application of formal models provides many benefits for the software and system development. 
In ``40 years of formal methods" \cite{bjorner201440}, Bj{\o}rner  and Havelund admit that the gap between academic research on formal methods and its integration in large industrial projects is yet to be bridged. 
There are a number of hindering factors for  adoption of formal methods in industry~\cite{enase2016cfm}. 
As crucial obstacles can be named 
lack of understandability and readability~\cite{Spichkova2013HFFM,spichkova2013we}, 
and our aim is to find appropriate ways to avoid these obstacles. 
Also, human factors play a crucial role and have to be taken into account~\cite{hffm_spichkova,spichkova2015human}.

Application of formal models  requires an interplay between formal and informal methods, which use different levels of formality in descriptions. 
A manual solution to this problem was suggested many years ago: 
Guiho and Hennebert reported a communication problem 
 in the SACEM project \cite{guiho1990sacem} between the verifiers and other
engineers, who were not familiar with the formal specification method. The problem was solved by providing
the engineers with a natural language description derived \emph{manually} from the formal
specification. For a large-scale projects, it would be too time-consuming to derive a natural language specification (NLS) manually. 
In this paper, we propose a framework for \emph{automated generation} of NLS from  the basic modelling artefacts, such as data type definitions, State Transition Diagrams (STDs), and architecture specifications. 

\textbf{Contributions:} 
The proposed solution would serve not only increasing the understandability of formal models, but also keeping the system documentation up-to-date.
System documentation is  an important part of the development process, but  
it is often considered by industry as a secondary appendage to the main part of the development -- modelling and implementation. It is hard to keep the documentation up-to-date if the system model is frequently changing during the modelling phase of the development. Thus, system requirements documents and the general systems description are not updated according to the system's or model's modifications. 
Sometimes the updates are overlooked, sometimes they are omitted on purpose. For example, it is because of timing or costs constraints on the project. As a result, the system documentation is often  outdated and does not describe the latest version of the system model. 
The question is whether we need to update the documentation  \emph{manually}, cf. \cite{spichkova2013we}.

\textbf{Outline:} 
The rest of the paper is organised as follows.
Section~\ref{sec:related} describes the related work.
Section~\ref{sec:framework} introduces the proposed framework and a 
a small case study  to illustrate the ideas of the framework. 
In Section~\ref{sec:conclusions} we summarise the paper and propose directions for future research.

%========================================
\section{Related work}
\label{sec:related}
 
The research field of automated translation from formal modelling languages to natural languages is almost uncovered, however, there are many approaches on automated generation of (semi-)formal specifications from natural language ones. 
Lee and Bryant \cite{lee2002automated} presented an approach automatically generate formal specifications in an object-oriented notation from NLS.
Cabral and Sampaio \cite{cabral2008formal} suggested to use a Controlled Natural Language (CNL), 
a subset of English to analyse system characteristics represented by a set of declarative sentences.   
CNL use restricted vocabulary, grammar rules in defined knowledge based for the aim of formal models generation. 
This also allows to generate structured models at different levels of abstraction, as well as provides formal refinement of user actions and system responses. 

Schwitter et al. \cite{schwitter2003ecole} introduced
ECOLE, an editor for a controlled language called PENG (Process-able English), that defines a mapping between English and First-Order Logic in order to verify requirements consistency, as well as to help  writing manuals and system specifications to improve documentation quality, which is our goal of generated specifications in natural language.

As several attempts have been made to automate the requirement capture, there is another approach for the automatic construction of Object-oriented design model in UML diagram from natural language requirement specification.  
Mala  and Uma  \cite{mala2006automatic} present a methodology that utilizes the automatic reference resolution and eliminates the user intervention.
The input problem statement is split into sentences for tagging by sentence splitter in order to get parts of speech for every word. 
The nouns and verbs are then identified by tagged texts based on simple phrasal grammars. 
Reference resolver is used to remove ambiguity by pronouns. 
The final text is then simplified by the normaliser for mapping the words into object-oriented system elements. 
The result produced by the system is compared with human output on the basic analysis of the text. 
The approach is promising to introduce a method to restructure the natural language text into modelling language in respect of system requirements specifications. 
Although there is a shortage of the efficiency in the tagger and reference resolver that result in unnatural expressions and misunderstandings, it can be improved by building a knowledge base for the system elements generation.

Juristo et al. \cite{juristo1999formal} introduced an approach to formalise the requirement analysis process. 
The goal of this approach was to generate conceptual models in a precise manner, which provides support for resolving difficulties of misunderstanding the system requirements.  
The approach is based on examining the information extraction at the beginning of the development process (i.e.,  describing the problems in natural language sentences),
and consists of two different activities: 
formalisation of the conceptual model  and creation of the formal model.  The limitation of this approach is in the difficulties to retrieve the rigorous and concise problem descriptions.

Gangopadhyay \cite{gangopadhyay2001conceptual} suggested to design a conceptual model from a functional model, expressed in natural language sentences. 
Although its application is mainly for  database applications, it can be extended to other  
design problems such as Web engineering and data warehousing. 
In order to interpret natural language expressions, 
Gangopadhyay applied the theory of Conceptual Dependencies developed by Schank, cf. \cite{schank1972conceptual}.
The main goal of this approach was to identify data elements from functional model expressed in NLS, 
to locate missing information, as well as to integrate all individual data elements into an overall conceptual schema for data model establishment. 
A prototype system using Oracle database management system has been implemented to contain a parser for information collection.  
However, the lexicon in use is developed incrementally and semi-automated, so domain specialists still need to manually categorise words and phrases, to ensure non-relevant words are included in the system during the development of the   conceptual 
model and to prevent systematic bias.

Bryant \cite{bryant2000object} suggested the theory of Two-Level Grammar for  natural language requirements specification, in conjunction with Specification Development Environment to allow user interaction to refine model concepts. This approach  allows the automation of the process of transition from requirements to design and implementation, as well as producing an understandable document on which software system will base on.  
 
Ilieva and Ormandjieva \cite{ilieva2005automatic} proposed an approach on transition of natural language software requirements specification into formal presentation.
The authors decided their method into three main processing parts: 
(1) the Linguistic Component as the text sentences to be analysed; 
(2) the Semantic Network as the formal NL presentation; and 
(3) modelling as the final phase of formal presentation of the specification. 
However, the approach  of Ilieva and Ormandjieva 
involves manual human analysis process, to break down problems into smaller parts that are easily understood.

%========================================
\section{Framework}
\label{sec:framework}

\begin{figure}[ht!]
  \centering
    \includegraphics[scale=0.5]{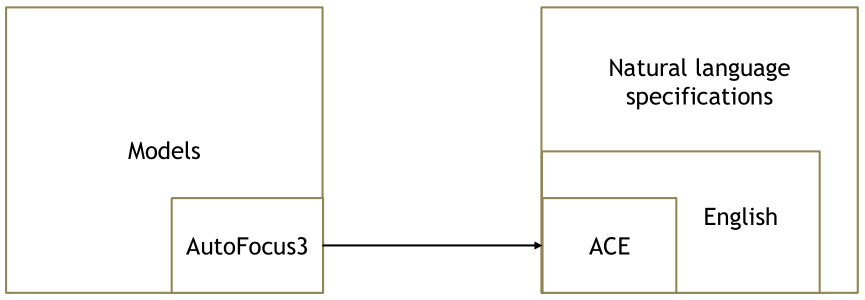} 
  \caption{Framework: Generation of natural language specifications from formal models}
  \label{fig:suggested_framework}
\end{figure}

Figure \ref{fig:suggested_framework} illustrates the general ideas of the suggested framework.
To build a prototype for generation of NLS from the basic modelling artefacts, we have selected the AutoFocus3 modelling tool  \cite{aravantinos2015autofocus,holzl201013} as the basis for our models, 
because this tool (1) embeds the basic modelling artefacts, (2) is open source, as well as (3) has a well defined formal syntax behind all its modelling elements.
 
AutoFocus3 is developed on system models based on the \Focus\ theory \cite{focustheory}  that allows to specify system on different levels of abstraction formally and precisely. 
Source code of AutoFocus3 models are coded in XML, which makes it easy to parse and to analyse. 
AutoFocus3 has many advantages and is constantly evolving through last 10 years. 
The tool was applied as a part of tool chain within a number of development methodologies,  
e.g., for safety-critical systems in general \cite{VerisoftXT_FMDS,verisoftxt_praxis,holzl2010safety}, and 
for automotive-systems~\cite{feilkas2011refined,feilkas2009top}. 
The tool can also be successfully applied for service-oriented modelling~\cite{broy2008service}, which gives us 
another reason to select AutoFocus3 for the framework we develop. 

To allow further formal analysis of the generated specification, we restrict English to its subset, Attempto Controlled English (ACE), cf. \cite{fuchs1996attempto}. 
Specifications written in ACE give the impression of being informal, though they are in fact formal and machine executable.  
ACE provides a set of principles and recommendations for the strategy: to reduce the amount of lexical resources and structural sentences for a specification text to be unambiguously represented, and to fulfil the communication gap between domain specialist and software developer. Basically, the construct of ACE specification is the declarative sentence that is expressive enough to allow both natural usage and computer-processed purpose \cite{fuchs2008attempto}. % 

 ~\\ 
\textbf{Implementation:} 
We are currently implementing an automated translator from the AutoFocus3 models to ACE sentences in the Python programming language. 
Python was chosen as the development language due to its rapid prototyping features,  
as well as due to its increasing uptake by researchers as a scientific software development language because of good code readability and maintainability.
 With regard to the Python performance, it is sufficient for many common tasks and turns out to be very close to C language for parsing a file and a tree-like structure, cf. \cite{sanner1999python}. 
 For the execution environment, we will research on the installation of ACE parsing engine, cf. \cite{aceparser}, to execute natural language sentences in Prolog, cf. \cite{swiprolog}.
 
 ~\\
\textbf{XML code of AutoFocus3 models.}
 While parsing the XML code of an AutoFocus3 model, we have to identify three core sections:
 \begin{itemize}
 \item Specifications of data types and functions/constants (introduced by the XML-tag \emph{rootElements} with the type \emph{Data Dictionary}, cf. below for an example from the SimpleTrafficLight case study).
 \item Specifications of the system and components architecture (introduced by the XML-tag  \emph{rootElements} with the type \emph{ComponentArchitecture});
 \item Specifications of the state machines, used to describe the behaviour of system components (introduced by the XML-tag \emph{containedElements} with the type \emph{StateAutomaton}):
 \end{itemize}
 As each of these parts consists of XML representation of the AutoFocus3 elements, we can define a translation schema for each of these elements to generate English sentences out of the XML code. The sentences should be conform to the ACE rules. 
 To validate that this constraint is fulfilled, we have to analyse syntax and semantics of the generated sentences.
 
 ~\\
\textbf{Translation schema.} 
Let us discuss the translation schema in more details, focusing for simplicity on the specifications of data types and functions/constants.
 The definition of each data type is provided within the XML-tag \emph{typeDefinitions}, where the keyword \emph{Enumeration} indicates that this is an enumeration type. The name of the data type is coded within the attribute \emph{name}. The elements of the type are introduces with the tag \emph{members}. 
		For the case of an enumeration type, we would have the following XML structure, 
		where $N$ is a natural number representing a number of elements in the data type, and  
		$i_1, \dots, i_{N+1}$ are some natural numbers representing internal  identifiers of AutoFocus3 elements:\\
	%	~
	%\\
	{\small 
	 \textsf{
	   $<$typeDefinitions xsi:type=\epar{org-fortiss-af3-expression-definitions:Enumeration} id=\epar{$i_1$} \\
	   name=\epar{TypeName}$>$\\
      $<$members id=\epar{$i_2$} name=\epar{MemberName$_1$} /$>$\\
      \dots\\
     $<$members id=\epar{$i_{N+1}$} name=\epar{MemberName$_N$} /$>$\\
    $<$/typeDefinitions$>$\\} }
    ~
	\\
	To generate an ACE sentence from this structure, we define two templates:
		\begin{itemize}
		\item
		For the case we have only one element, i.e., $N=1$, we would use the template
	
		\textsf{TypeName} is a datatype. It consists-of  one element that is \textsf{MemberName$_1$}.
		\item
		For the case we have more than one element, i.e., $N>1$, we would use the template
		
		\textsf{TypeName} is a datatype. It consists-of  $N$ elements that are \textsf{MemberName$_1$}, \dots, \textsf{MemberName$_N$}.
		\end{itemize}
~\\  	 
%	 \newpage
%	 \noindent
	The definition of each function/constant  is provided within the tag \emph{function}, where its name and value are coded within the attributes \emph{name} and \emph{value}.  
	For the case of constant function, we would have the following XML structure, where $j_1, j_2$ are some natural numbers representing internal  identifiers of AutoFocus3 elements:
	\\ ~
	\\
	{\small 
	 \textsf{
 	$<$functions id=\epar{$j_1$} $>$ \\
      ~~~ $<$function id=\epar{$j_2$} name=\epar{ConstantName} /$>$\\
      ~~~ $<$definition$>$\\
      ~~~ $<$statements xsi:type=\epar{org-fortiss-af3-expression-terms-imperative:Return}$>$\\
      ~~~ $<$value xsi:type=\epar{org-fortiss-af3-expression-terms:IntConst} value=\epar{ConstantVaue}/$>$\\
      ~~~ $<$/statements$>$\\
      ~~~ $<$/definition$>$\\
      ~~~ $<$returnType xsi:type=\epar{org-fortiss-af3-expression-types:TInt} /$>$\\
    $<$/functions$>$\\
	 } }
	  ~\\ 
	To generate an ACE sentence from this structure, we define the following template:\\
	
	\textsf{ConstantName} is a constant. It is equal to \textsf{ConstantVaue}.

 ~\\
Similar translation patterns apply for architecture specifications  and state transition diagram sections.

%========== 
 ~\\
\textbf{ACE: Syntax check.} 
ACE supports declarative sentences, which includes simple sentences, there is/are-sentences, boolean formulas, composite sentences, interrogative sentences, imperative sentences.
ACE construction rules determine whether an English sentence is an ACE sentence, cf.~\cite{aceconstrules}. Each ACE sentence is an acceptable English sentence, but not every English sentence is justified as a valid ACE sentence.  
Thus, to be conformed to ACE construction rules, an NLS in English should be constructed from the following elements:
\begin{itemize}
\item Function words: determiners, quantifiers, coordinators, negation words, pronouns, query words, modal auxiliaries, \epar{be}, Saxon genitive marker's;
\item Fixed phrases: \epar{there is}, \epar{it is true that};
\item Content words: nouns, verbs, adjectives, adverbs, prepositions.
\end{itemize}
The function words and fixed phrases are predefined and cannot be changed, 
whereas content words can be modified by users within the lexicon format, cf.~\cite{acelexicon}. 
The content words cannot contain blank spaces. For instance, \epar{interested in} should be reformulated to \epar{interested-in}.

~\\
\textbf{ACE: Semantics check.}  
The mentioned above rules cannot remove all ambiguities in English. 
To avoid ambiguity, ACE provides a set of interpretation rules.  
Thus, each ACE sentence can have only one meaning, based on its syntax and on syntax of previous sentences.

The correctness  of the generated sentences can be validated by the ACE query sentences,
 cf. \cite{fuchs2008attempto}.  
 They can be subdivided into three forms that are $yes/no$-questions (questions that require answer \epar{yes} or \epar{no} ), $wh$-questions (questions starting with the words \epar{What}, \epar{When}, \epar{Where}, etc.), and \emph{how much/many}-questions, cf. \cite{aceconstrules}.
For example, we could use the following questions to check the definition of an enumeration data type $XDataType$:
\begin{itemize}
\item
What is  $XDataType$?  
 \item
How many elements does $XDataType$ have? 
 \item
Is $SomeElementName$ an element of $XDataType$? 
\end{itemize}

\begin{figure}[ht!]
  \centering
	\includegraphics[width=\textwidth]{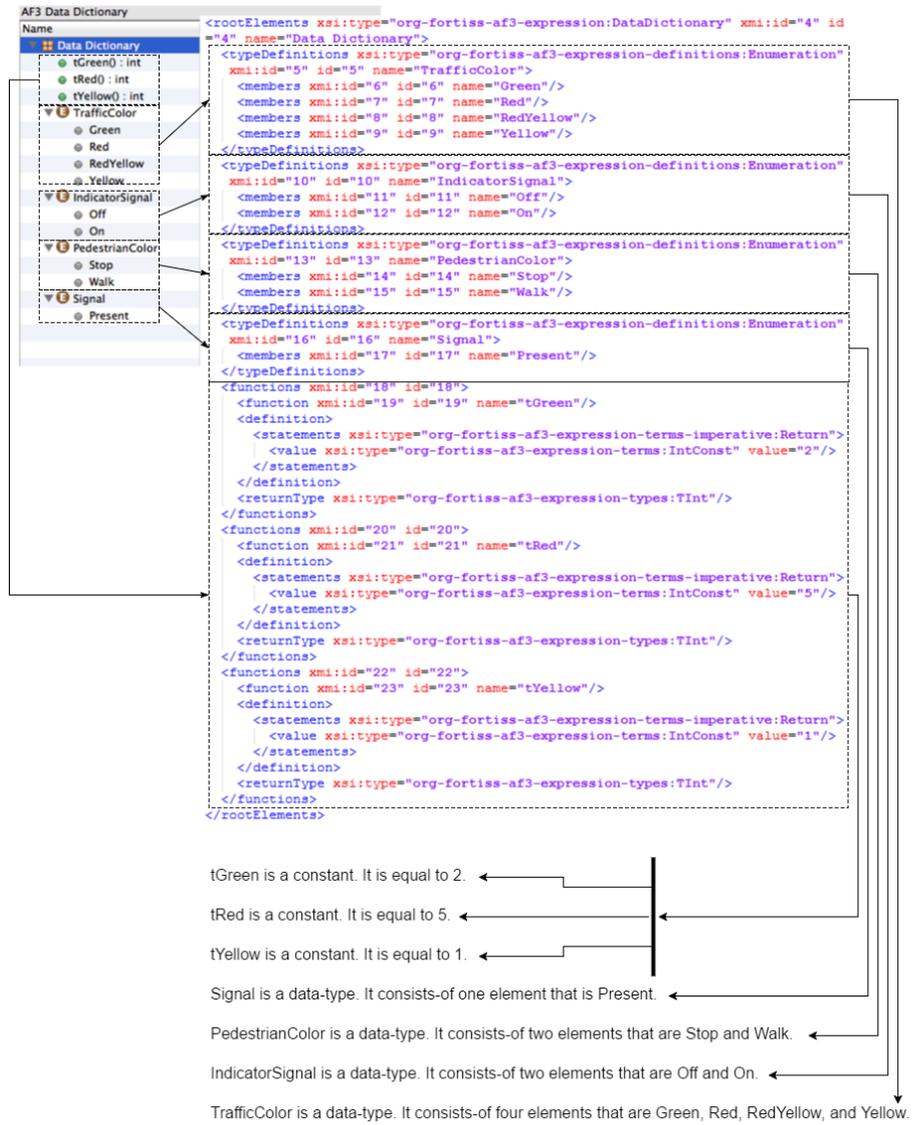}
  \caption{Mapping from AutoFocus 3 data types to ACE sentences}
  \label{fig:mapping_af3_xml_ace_datatypes}
\end{figure}

\begin{figure}[ht!]
  \centering
    \includegraphics[width=0.7\textwidth]{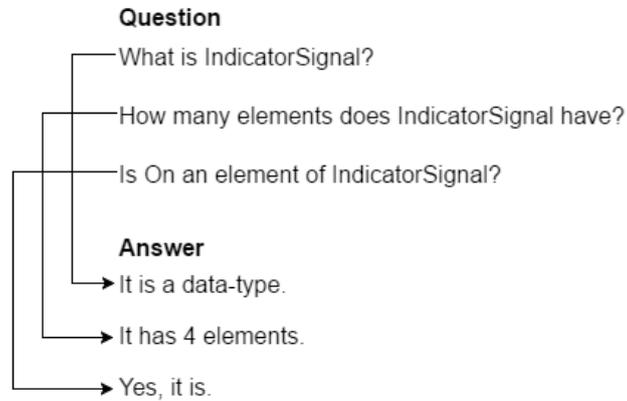}
  \caption{Validation the generated sentences using ACE-questions}
  \label{fig:ACEquestionsTL}
\end{figure}

%============================================
~\\
\textbf{Case study: SimpleTrafficLight system.}
We present the core ideas of the framework on example of a small case study, Simple Traffic Lights, 
 introduced  by Lam and  Teufl in  \cite{simpletraffic}.  
In the Simple Traffic Lights  case study, we the following elements in the data definitions section:
\begin{itemize}
\item 
Functions \emph{tGreen}, \emph{tRed}, and \emph{tYellow}  
that return a constant integer value to represent the time in seconds for the active pedestrian or traffic light.
\item 
Enumeration data types:  
	\begin{itemize}
	\item 
	\emph{pedastrianColor}: pedestrian lights (\emph{Stop}, \emph{Walk});
	\item
	\emph{TrafficColor}: traffic lights (\emph{Green}, \emph{Red}, \emph{RedYellow}, \emph{Yellow});
	\item
	\emph{Signal}: one-element data type to represent the \emph{Present} signal;
	\item
	\emph{IndicatorSignal}: pedestrian requests to pass the street (\emph{Off}, \emph{On}).
	\end{itemize}
\end{itemize}
Figure~\ref{fig:mapping_af3_xml_ace_datatypes} illustrates the translation process from the AutoFocus3 data types and the corresponding XML descriptions, to ACE sentences. 
After translation, we check the definition of each data type as shown on Table~\ref{tab:ACEquestionsTL} and in Figure~\ref{fig:ACEquestionsTL}.

In a similar manner the natural language description of the system and components architecture as well as of state machines,
 representing components behaviour, are generated and checked.

\begin{table}
\begin{center}
\caption{Validation the generated sentences using ACE-questions}
\label{tab:ACEquestionsTL}
\begin{tabular}{ |l|l|} 
 \hline
Question		&		Answer	\\ [0.5ex]
 \hline
 \hline
What is IndicatorSignal?  		& 	 It is a data-type. \\ [0.5ex]
 \hline
How many elements does IndicatorSignal have? 		&	It has 4 elements.	\\ [0.5ex]
 \hline
Is On an element of IndicatorSignal? 			&	Yes, it is.	\\ [0.5ex]
 \hline
\end{tabular}
\end{center}
\end{table}

\newpage 

\newpage 
~

%============================================
\section{Conclusions and Future Work}
\label{sec:conclusions}

This paper introduces our ongoing work on  NLS from formal models. 
The goal of our current work is to generate documentation in English from the basic modelling artefacts of the AutoFocus3 modelling language, 
that are data types, state machines, and architectural components. 
This would allow to have an easy-to-read and easy-to-understand specifications of systems-under-development, written in English. 
To allow further formal analysis of the generated specification, we restrict English to its subset, ACE. 
The proposed framework, in its current version, can be applied to  build a prototype for generation of ACE specifications from the  AutoFocus3 models.

The future work focuses on the implementation of an prototype translator from AutoFocus3 to ACE, as well as on the extension of the framework to other formal modelling languages.

\bibliographystyle{abbrv}

\end{document}